# A re-interpretation of the concept of mass and of the relativistic mass-energy relation

**Stefano Re Fiorentin**[1]

**Summary**. For over a century the definitions of mass and derivations of its relation with energy continue to be elaborated, demonstrating that the concept of mass is still not satisfactorily understood. The aim of this study is to show that, starting from the properties of Minkowski spacetime and from the principle of least action, energy expresses the property of inertia of a body. This implies that inertial mass can only be the object of a definition – the so called mass-energy relation - aimed at measuring energy in different units, more suitable to describe the huge amount of it enclosed in what we call the "rest-energy" of a body. Likewise, the concept of gravitational mass becomes unnecessary, being replaceable by energy, thus making the weak equivalence principle intrinsically verified. In dealing with mass, a new unit of measurement is foretold for it, which relies on the de Broglie frequency of atoms, the value of which can today be measured with an accuracy of a few parts in $10^9$.

**Keywords**  Classical Force Fields; Special Relativity; Classical General Relativity; Units and Standards

## 1. Introduction

Albert Einstein in the conclusions of his 1905 paper "*Ist die Trägheit eines Körpers von seinem Energieinhalt abhängig?*" ("*Does the inertia of a body depend upon its energy content?*") [1] says "*The mass of a body is a measure of its energy-content*". Despite of some claims that the reasoning that brought Einstein to the mass-energy relation was somehow approximated or even defective (starting from Max Planck in 1907 [2], to Herbert E. Ives in 1952 [3], to Mendel Sachs in 1973 [4] and to John Stachel and Roberto Torretti in 1982 [5]), and in spite of the fact that Einstein himself does not seem to have been satisfied with his 1905 derivation or with any other of his various derivations, it is generally recognized that "*the mass-energy relation, usually symbolized by $E = mc^2$, is one of the most important and empirically best confirmed statements in physics*", to cite the words of Max Jammer, author of probably one of the most comprehensive studies on the concept of mass [6], [7]. The link established between energy and mass, however, has never led to the abolition of the concept of mass, which has always been considered to be an inherent and independent property of a body. Disputes on the definition of mass, and derivations of the mass-energy relation based on fewer and fewer assumptions, continued throughout last century and are still continuing.

Among definitions of mass, it is worth mentioning those of C.G. Pendse in 1937 [8], V.V. Narlikar in 1938 [9], John Charles Chenoweth McKinsey and collaborators in 1953 [10], Rudolf Carnap in 1966 [11], Arnold Koslow in 1968 [12], Goodinson and Luffman in 1985 [13], Andreas Kamlah in 1988 [14], Hans Jürgen Schmidt in 1993 [15], Adonai Schlup Sant'Anna in 1996 [16] and, most recently, Enzo Zanchini and Antonio Barletta in 2008 [17].

Likewise, new derivations of the mass-energy relation have continued to be developed. Among most recent ones, those of Daniel J. Steck and Frank Rioux in 1983 [18], Mitchell J. Feigenbaum and N. David Mermin in 1988 [19], Fritz Rohrlich in 1990 [20], Ved Prakash Srivastava [21] and Ralph Baierlein [22] in 1991.

---

[1]  Fiat Group Automobiles, Advanced Engineering Department, 200 Corso Agnelli, 10135 Turin, Italy
 e-mail: stefano.refiorentin@fiat.com

With two distinct definitions for mass and energy, the mass-energy relation has always been regarded as a factual law, stating the equivalence of the two quantities. The intention here is to show that, under the axiomatic basis of both Minkowski spacetime properties and the principle of least action, the concept of energy can be naturally derived as well as its role of inertia of a body; in this context the mass simply becomes the object of a definition aimed at describing energy in a different unit of measurement: $m := E/c^2$.

The mass referred to here is the inertial mass, representing a measure of an object's resistance to changing its state of motion when a force is applied. As is well known, another type of mass is used in Physics: the gravitational mass, responsible of the weight of a body. Einstein developed his general theory of relativity starting from the assumption that a coincidence exists between the inertial and the gravitational mass, and that no experiment will ever detect a difference between them (the weak version of the equivalence principle). However, in the words of Rindler [23]: "*the equality of inertial and active gravitational mass [...] remains as puzzling as ever*". In the same way in which the concept of inertial mass can be considered pleonastic, also the concept of gravitational mass can be fully replaced by energy. This inherently makes the equivalence principle fulfilled.

## 2. Momentum and energy of a particle and of a system of particles

We start considering a particle freely moving in an inertial frame of reference in the context of *Minkowski spacetime*, satisfying special relativity postulates together with the properties of homogeneity and isotropy. According to the *principle of least action* [24], in correspondence of the actual motion of the particle there exist a certain invariant integral *S*, called *action*, which gets its minimum value. The action can be represented as an integral with respect to the time

$$S = \int_{t_1}^{t_2} L(\mathbf{x}(t), \mathbf{u}(t), t) dt \qquad (1)$$

The coefficient *L* of *dt* represents the *Lagrange function* of the system, while $\mathbf{x}(t)$ denotes the spatial coordinates of the particle and $\mathbf{u}(t)$ is its velocity.[2] The minimum condition for *S* on the actual path implies that function *L* satisfies the *Euler-Lagrange equations*

$$\frac{d}{dt}\left(\frac{\partial L}{\partial \mathbf{u}}\right) - \frac{\partial L}{\partial \mathbf{x}} = 0. \qquad (2)$$

Thanks to the homogeneity of spacetime, the Lagrange function of the freely moving particle cannot explicitly depend on either space or time. In particular, homogeneity of space, in view of equations (2), implies

$$\frac{\partial L}{\partial \mathbf{x}} = \frac{d}{dt}\left(\frac{\partial L}{\partial \mathbf{u}}\right) = 0 \qquad (3)$$

It can be concluded that, as a direct consequence of the homogeneity of space and in agreement with Noether's Theorem [25], the vector quantity

$$\mathbf{p} := \frac{\partial L}{\partial \mathbf{u}} \qquad (4)$$

remains constant during the motion of the free particle; $\mathbf{p}$ is called *momentum of the particle*. The homogeneity of time, in its turn, leads to

$$\frac{dL}{dt} = \frac{\partial L}{\partial \boldsymbol{x}} \cdot \boldsymbol{u} + \frac{\partial L}{\partial \boldsymbol{u}} \cdot \frac{d\boldsymbol{u}}{dt}. \tag{5}$$

that, using equations (2) to replace $\partial L/\partial \boldsymbol{x}$ irrespectively of homogeneity of space, can be put in the form:

$$\frac{dL}{dt} = \frac{d}{dt}\left(\frac{\partial L}{\partial \boldsymbol{u}}\right) \cdot \boldsymbol{u} + \frac{\partial L}{\partial \boldsymbol{u}} \cdot \frac{d\boldsymbol{u}}{dt} = \frac{d}{dt}\left(\frac{\partial L}{\partial \boldsymbol{u}} \cdot \boldsymbol{u}\right), \tag{6}$$

or

$$\frac{d}{dt}\left(\boldsymbol{u} \cdot \frac{\partial L}{\partial \boldsymbol{u}} - L\right) = 0. \tag{7}$$

---

[2] In Appendix a derivation is reported of the principle of least action from quantum mechanics postulates.

Therefore we find that the quantity that appears into brackets in equation (7) is an integral of motion. This quantity is called *energy* of the particle and the following definition is adopted

$$E := \left(\boldsymbol{u} \cdot \frac{\partial L}{\partial \boldsymbol{u}} - L\right). \tag{8}$$

This definition can be assigned a general validity, extending it to cover also the cases in which the particle is not free and energy is therefore not a constant of motion.

Let us now take into account a closed system of $n$ distinct interacting particles, described by the Lagrange function

$$L(\boldsymbol{x}_1, \boldsymbol{x}_2, ..., \boldsymbol{x}_n; \boldsymbol{u}_1, \boldsymbol{u}_2, ..., \boldsymbol{u}_n), \tag{9}$$

depending on the coordinates $\boldsymbol{x}_i$ and on the velocities $\boldsymbol{u}_i$ of the single particles but not directly on time by virtue of its homogeneity. The principle of least action, in this case, gives the $n$ Euler-Lagrange equations

$$\frac{d}{dt}\left(\frac{\partial L}{\partial \boldsymbol{u}_i}\right) - \frac{\partial L}{\partial \boldsymbol{x}_i} = 0 \qquad (i = 1, 2, ..., n). \tag{10}$$

Thanks to the homogeneity of space the mechanical properties of a closed system do not vary as a consequence of a parallel translation in space of the whole system. Let us therefore consider an infinitesimal translation $\boldsymbol{\varepsilon}$ and require the Lagrange function (9) not to vary. The variation of $L$ as a consequence of the translation, velocities remaining unchanged, is given by

$$\delta L = \sum_{i=1}^{n} \frac{\partial L}{\partial \boldsymbol{x}_i} \delta \boldsymbol{x}_i = \boldsymbol{\varepsilon} \sum_{i=1}^{n} \frac{\partial L}{\partial \boldsymbol{x}_i}. \tag{11}$$

Due to the arbitrariness of $\boldsymbol{\varepsilon}$, the condition $\delta L = 0$ is equivalent to

$$\sum_i \frac{\partial L}{\partial \boldsymbol{x}_i} = 0 \tag{12}$$

and, by virtue of Euler-Lagrange equations (10)

$$\sum_i \frac{d}{dt}\left(\frac{\partial L}{\partial \boldsymbol{u}_i}\right) = \frac{d}{dt}\left(\sum_i \frac{\partial L}{\partial \boldsymbol{u}_i}\right) = 0. \tag{13}$$

It can be concluded that the vector quantity

$$\boldsymbol{p} := \sum_i \frac{\partial L}{\partial \boldsymbol{u}_i} \tag{14}$$

remains constant during the evolution of the system; $p$ is called *momentum of the system of particles*, while the quantity

$$p_i := \frac{\partial L}{\partial u_i} \tag{15}$$

in agreement with definition (4), is called *momentum of the i-th particle*.
Thanks to the homogeneity of time, and in analogy with the case of a single particle, we get

$$\frac{d}{dt}\left(\sum_i u_i \cdot \frac{\partial L}{\partial u_i} - L\right) = 0. \tag{16}$$

The quantity

$$E := \sum_i u_i \cdot \frac{\partial L}{\partial u_i} - L \tag{17}$$

that remains constant during the evolution of the system, is called *total energy of the system of particles*.

## 3. Force acting on a particle

In terms of the momentum $p_i$, equations (10) become

$$\frac{\partial L}{\partial x_i} = \frac{dp_i}{dt}. \tag{18}$$

The quantity $\partial L/\partial x_i$ represents therefore the rate of change of particle momentum: it has been given the name of *force acting on the i-th particle*

$$f_i := \frac{\partial L}{\partial x_i}. \tag{19}$$

Using definition (19), equation (18) can be put in the form

$$f_i = \frac{dp_i}{dt}, \tag{20}$$

which is essentially the expression of Newton's second law, while condition (12) represents Newton's third law.

## 4. The expressions for the energy and momentum of a free particle

In order to obtain explicit expressions for the energy and the momentum of a particle, it is necessary to know its Lagrange function. We start by deriving it in the case of a freely moving particle. We saw previously that homogeneity of spacetime implies that $L$ cannot explicitly depend on either $x(t)$ or time. Isotropy of space, in turn, forces $L$ to depend only on the modulus of $u$, say $u^2$

$$L = L(u^2). \tag{21}$$

Since, as already stated, action is an invariant with respect to Lorentz transformations[3], we deduce that $Ldt$ is invariant. For a free particle, the only non trivial invariant is its proper time element $d\tau$. Therefore it can be argued that

$$L(u^2)dt = -\alpha\, d\tau = -\alpha\sqrt{1-[u(t)/c]^2}\,dt, \qquad (22)$$

where $\alpha$ is a positive constant, and the "minus" sign guarantees that action $S$ has a minimum, since integral (1) gets its maximum value along the world line connecting the starting point 1 to the end point 2. From definition (8) and expression (22), the energy of the free particle can be derived

$$E := \mathbf{u}\cdot\frac{\partial L}{\partial \mathbf{u}} - L = \frac{\alpha}{\sqrt{1-(u/c)^2}}. \qquad (23)$$

It can be observed that when the particle velocity is zero (or when we are in the particle frame of reference) its energy does not go to zero but takes the value

$$E_0 := \lim_{u\to 0} E = \alpha, \qquad (24)$$

generally referred to as the *rest energy* of the particle. We can therefore write the energy in the form

$$E = \frac{E_0}{\sqrt{1-(u/c)^2}}, \qquad (25)$$

while the Lagrange function of the free particle becomes

$$L = -E_0\sqrt{1-(u/c)^2}. \qquad (26)$$

The *kinetic energy* of the particle is defined as the difference between its energy $E$ and its rest energy $E_0$

$$E_k := \frac{E_0}{\sqrt{1-(u/c)^2}} - E_0. \qquad (27)$$

At speeds that are small compared to $c$, the following approximation can be adopted for the kinetic energy

$$E_k = E_0\left[\frac{1}{2}\left(\frac{u}{c}\right)^2 + \frac{3}{8}\left(\frac{u}{c}\right)^4 + \ldots\right] \approx \frac{1}{2}E_0\left(\frac{u}{c}\right)^2. \qquad (28)$$

This expression highlights that, $(u/c)^2$ being small by assumption, the rest energy $E_0$ of macroscopic objects must necessarily be a large number, if measured in the same unit of measurement of $E_k$. This fact raises the question regarding whether it is reasonable to measure both $E$ and $E_0$ with the same unit of measurement, a question to which we will return in the following paragraph.

Moving from the case of a single free particle to a system of interacting particles, in building the Lagrange function we have to sum terms analogous to (26), one for each particle, plus a term depending on particle coordinates, which accounts for the interactions among particles and/or between particles and fields that permeate the whole space

---

[3] The reason why action is Lorentz-invariant can be ascribed to the invariance of phase, in view of the link between action and phase shown in appendix

$$L = -\sum_{i=1}^{n} E_{0i} \sqrt{1-(u_i/c)^2} - U(\mathbf{x}_1, \mathbf{x}_2, ..., \mathbf{x}_n). \tag{29}$$

Using this expression into (15), we get the following explicit expression for $\mathbf{p}_i$

$$\mathbf{p}_i = \frac{1}{c^2} \frac{E_{0i}}{\sqrt{1-(u_i/c)^2}} \mathbf{u}_i = \frac{E_i}{c^2} \mathbf{u}_i. \tag{30}$$

## 5. Inertia and the definition of inertial mass

Taking into account that momentum $\mathbf{p}_i$ is given by expression (30), by differentiation with respect to time and substitution into (20) we get

$$\mathbf{f}_i = \frac{d\mathbf{p}_i}{dt} = \frac{E_i}{c^2} \frac{d\mathbf{u}_i}{dt} + \frac{dE_i}{dt} \frac{\mathbf{u}_i}{c^2}. \tag{31}$$

The expression of $dE_i/dt$ can be obtained by differentiation of expression (25) referred to the *i-th* particle, under the assumption that $E_{0i}$ is constant (the force is a purely mechanical one)

$$\frac{dE_i}{dt} = \frac{E_i}{c^2 - u_i^2} \left( \mathbf{u}_i \cdot \frac{d\mathbf{u}_i}{dt} \right). \tag{32}$$

Defining the three-acceleration of the *i-th* particle in the reference frame as

$$\mathbf{a}_i := \frac{d\mathbf{u}_i}{dt}, \tag{33}$$

and substituting (32) and (33) into (31), we obtain:

$$\mathbf{f}_i = \frac{E_i}{c^2} \left[ \mathbf{a}_i + (\mathbf{u}_i/c) \frac{\mathbf{a}_i \cdot (\mathbf{u}_i/c)}{1-(u_i/c)^2} \right]. \tag{34}$$

Taking into account the components of the acceleration $\mathbf{a}_i$ parallel ($\mathbf{a}_{i\parallel}$) and orthogonal ($\mathbf{a}_{i\perp}$) to the velocity vector $\mathbf{u}_i$, we can re-write equation (34) in the form

$$\mathbf{f}_i = \frac{E_i}{c^2} \left[ \mathbf{a}_{i\perp} + \frac{1}{1-(u_i/c)^2} \mathbf{a}_{i\parallel} \right]. \tag{35}$$

The coefficient of proportionality between force and acceleration accounts for the *inertia* of the body, that is, its resistance to the change of its state of motion when a force is applied to it. The quantity representing the inertia of a body is given the name of *inertial mass*. On the basis of expression (35) we would be led to conclude that the inertial mass of a particle is not a scalar quantity, having different components in the direction of particle velocity and in the orthogonal directions. However, we have to recognize that in the four-dimensional spacetime the concept of acceleration does not coincide with the one characteristic of ordinary three-space. In fact, the covariant expression of the acceleration, given by the four-vector that one gets by differentiating twice the spacetime particle position four-vector $X_i^\mu = \{ct, \mathbf{x}_i\}$ with respect to proper time

$$A_i{}^\mu := \frac{d}{d\tau}\left(\frac{dX_i{}^\mu}{d\tau}\right), \tag{36}$$

has components, expressed in terms of the particle velocity $u_i$ and the three-acceleration $a_i$, given by

$$A_i{}^\mu = \frac{1}{1-(u_i/c)^2}\left\{\frac{a_i \cdot (u_i/c)}{1-(u_i/c)^2},\ a_i + (u_i/c)\frac{a_i \cdot (u_i/c)}{1-(u_i/c)^2}\right\}. \tag{37}$$

Comparing the spatial component of this four-vector with the expression in square brackets in equation (34), we deduce that this last expression has to be used in place of the three-acceleration $a_i$ to properly describe the relativistic acceleration of the particle.

As a consequence, equation (34) shows that the energy $E_i$ of the particle, apart from the constant $c^2$ (that appears by virtue of the fact that we do not measure time in the units $ct$), embodies its inertia and therefore can be regarded as the particle inertial mass. We are led to the definitions

$$m_i := E_i/c^2 \tag{38}$$

and

$$m_{0_i} := E_{0_i}/c^2, \tag{39}$$

where $m_{0_i}$ is called the *rest inertial mass* of the *i-th* particle.

First of all, we have to conclude that the concept of inertia of a body is a direct consequence of the principle of least action and of Minkowski spacetime properties and not something that we have to postulate separately, as happens in Newtonian mechanics. If, nonetheless, the nature of inertia looks mysterious, we have to recognize that it will be thus as far as the a-priori assumptions that stay at the origin of the principle of least action[4] will be such and it is non possible to explain it on the basis of more fundamental concepts. In any case, inertia needs not to be justified by exogenous phenomena, like the effect of distant masses in what Albert Einstein called Mach's Principle [26].

Definitions (38) and (39) provide a new interpretation of the mass-energy relation, which ceases to be regarded as a law of nature connecting two distinct quantities (mass and energy), but instead becomes simply the definition of inertial mass. This last is simply another way of representing and measuring the energy of a body, with the advantage of a unit of measurement more adequate than that of energy to express the high energy content of $E_0$ that characterizes macroscopic bodies. In fact, the unit-of-mass, whatever it is, corresponds exactly to $c^2$ units-of-energy.

As concerns units of measurement, there are in principle two possibilities: define the unit of energy to become a fundamental unit of measurement and use definition (38) to get the unit of mass, or, vice versa, first define the unit of mass as a fundamental unit and then get the unit of energy. This second option is preferred, not only for historical reasons, but also because masses can be measured and compared with much more accuracy than energies. If we combine definitions (8) and (38), a definition of mass that does not need the prior definition of energy is obtained:

$$m := \frac{1}{c^2}\left(u \cdot \frac{\partial L}{\partial u} - L\right) \tag{40}$$

In the literature the question of what, precisely, is the conceptual meaning of the equation $E = mc^2$ has been raised since its origin. According to Jammer [7], "*at least two different interpretations have been proposed in the literature on this subject. According to one interpretation the relation expresses the convertibility of mass into energy or inversely of energy into mass, with*

---

[4] See appendix

*one entity being annihilated and the other being created. According to another interpretation the equation expresses merely a proportionality between two attributes or manifestations of one and the same ontological substratum without the occurrence of any annihilative or creative process.*" Definitions (38) and (39) rule out any ambiguity: inertial mass and energy not only represent the same property of a body, but inertial mass can be regarded as a conceptually unnecessary entity, were not for its practical usefulness in describing the huge amounts of energy embodied into the rest energy of macroscopic objects. It is also worthwhile to underline, as does Ernest F. Baker [27], that *mass* must not be confused with *matter*: all matter has the property of mass, but not all mass has the property of matter; mass is always conserved, matter not. We identify matter with rest energy: both the energy content and the inertial properties of matter are described by its mass.

## 6. Gravitational mass

The previous analysis has dealt only with the inertial mass of a body. Nevertheless it is well known that a body is also attributed a gravitational mass, conceptually distinct from the inertial one, and responsible of the weight of the body. It may be asked therefore what is the relationship of gravitational mass with energy and therefore with inertial mass. When speaking of gravitational mass, one has to distinguish between the *active gravitational mass* $m_a$, which specifies the body's role as the source of a gravitational field, and the *passive gravitational mass* $m_p$, which specifies the body's susceptibility to being affected by this field. It can be shown (e.g. see [7] pp. 93-94) that by assuming Newton's third law or, equivalently, the conservation of momentum, then the active and passive gravitational masses of every body, though conceptually different, are numerically equal. The relationship between $m_i$ and $m_p$ is governed by the so called *weak principle of equivalence* [28], which states that the ratio $m_p/m_i$ is the same for all bodies, regardless of their weight, size, shape, structure or material composition, or, in appropriate units, $m_i = m_p$. Thanks to this principle, Einstein realized that it was possible to "transform away" a homogeneous gravitational field locally and thus to extend the applicability of special relativity to the case of uniformly accelerated reference frames. Therefore it can be asserted that general relativity owes its inception – apart from its methodological postulate of the general covariance of the physical laws – to the proportionality between $m_i$ and $m_p$. The principle of equivalence has never been successfully explained on the basis of more fundamental aspects, and it is often said that it was explained by Einstein himself in his general theory of relativity.

Now, having proved that inertial mass is but another way of representing energy, there is no longer need to refer to the gravitational mass, but we can directly state that energy itself has the additional property of generating the gravitational field, described by the Einstein equations of general relativity. In other words, we can substitute the postulate that energy determines the space-time curvature tensor for the weak equivalence principle.

## 7. Conclusions

Jammer says (see [7] pag. 144) "*If it were possible to define the mass of a body or particle on its own in purely kinematical terms and without any implicit reference to a unit of mass, such a definition might be expected to throw some light on the nature of mass*". Under present axiomatic basis, mass is inherently another way of measuring energy, representing a step in this direction: the problem is shifted to the explanation of the nature of rest-energy. This last problem is tackled, within the so called Standard Model, by the Higgs mechanism, developed in 1964 by Peter Higgs [29]. It is based on the assumption of the existence of a scalar field, the "Higgs field", which permeates all space. By coupling with this field, a zero rest-energy particle acquires a certain amount of potential energy and therefore a non-null rest-energy, that we interpret as mass. The stronger the coupling, the more massive the particle.

**Appendix**

The principle of least action can be explained on the basis of the postulates of quantum mechanics that, following Feynman [30], we express in the form

Postulate n.1 -  Any fundamental event is governed by a probability of occurrence given by the square of a complex "probability amplitude".

Postulate n.2 -  The probability amplitude for some event is given by adding together the contributions of all possible alternatives which may lead to the same event.

Postulate n.3 -  To each particle is associated a characteristic scalar quantity, which we indicate with ν, depending, in general, on its space-time coordinates and its velocity, such that the phase $\varphi$ of the probability amplitude associated to a particular propagation path (or alternative) is given by the law

$$\varphi = 2\pi \int_{t_1}^{t_2} \nu(\boldsymbol{x}(t), \boldsymbol{u}(t), t)\, dt, \tag{1a}$$

In view of its role, the quantity ν is referred to as "frequency" of the particle (or "de Broglie frequency", after Louis-Victor-Pierre-Raymond, 7th duke de Broglie, who was the first to postulate it in 1924 [31]).

In order to find the overall probability amplitude for a given process, then, one adds up, or integrates, the probability amplitudes over the space of all possible histories of the system in between the initial and final states, including histories that are absurd by classical standards. In calculating the amplitude for a single particle to go from point 1 to point 2 in a given time, it is correct to also include histories in which the particle describes elaborate paths. The integral (generally called *path integral*) assigns all of these histories probability amplitudes of equal magnitude but with varying phase, determined by equation (1a). When frequency is very large (as it is in the case of macroscopic objects), the path integral is dominated by solutions which are in the neighbourhood of stationary points of phase, since here the amplitudes of similar histories tend to constructively interfere with one another. Conversely, for paths that are far from the stationary points of (1a), the phase varies rapidly for similar paths, and amplitudes tend to cancel. Therefore the important parts of the path integral—the significant possibilities—in the limit of large frequency, simply consist of solutions of the equation

$$\delta \int_{t_1}^{t_2} \nu(\boldsymbol{u}(t), \boldsymbol{x}(t), t)\, dt = 0. \tag{2a}$$

We therefore deduce that the Lagrange function of a particle has to be proportional to its frequency

$$L(\boldsymbol{x}(t), \boldsymbol{u}(t), t) := h\nu(\boldsymbol{x}(t), \boldsymbol{u}(t), t). \tag{3a}$$

The constant $h$ is called *Planck constant*, since it was first introduced by Max Karl Ernst Ludwig Planck [32]. Its numerical value depends on the units of measurement chosen for energy and time. Irrespectively of the choice of the units of measurement, $h$ (like $c$) is a fundamental constant of nature, representing, according to definition (8), the energy of a particle that, viewed from its rest frame ($\boldsymbol{u}=0$), has a frequency of one [unit-of-time$^{-1}$].

The unit of time under the *Système International d'Unités* is a fundamental unit of measurement, the second, defined as the duration of 9,192,631,770 periods of the radiation corresponding to the transition between the two hyperfine levels of the ground state of the caesium 133 atom. On the contrary, energy in the SI is a derived unit, based on the units of mass, length and time. Present definition of the unit of mass is based on the international prototype of the kilogram, an artefact dating back to the 1880s. At present there is considerable effort worldwide aimed at replacing this artefact definition by one based on physical constants. On the basis of (3a) we have the opportunity to define a useful unit of measurement for mass assigning $h$ an arbitrary but appropriate value, just as the metre is now defined from the second by specifying the vacuum speed of light as exactly $299\,792\,458\,[\mathrm{ms}^{-1}]$. To be compliant with present unit of mass, the kilogram, we propose to assign $h$ the exact value:

$$h = 6.626\,068\,96 \times 10^{-34} [\mathrm{kg} \cdot \mathrm{m}^2 \cdot \mathrm{s}^{-1}] \tag{4a}$$

on the basis of most recent CODATA values [33]. Definition (40), rewritten in the form

$$m := \frac{h}{c^2}\left(\boldsymbol{u} \cdot \frac{\partial \nu}{\partial \boldsymbol{u}} - \nu\right) \tag{5a}$$

by virtue of (3a), makes the kilogram correspond to the inertia of a body whose de Broglie frequency, as viewed in its rest frame, is exactly $1.35639273325163 \times 10^{50}$ Hz. "*This is an awkwardly large number*" - says J.W.G. Wignall [34], who proposes a definition of mass based essentially on the same approach - "*but this is of little concern because macroscopic mass measurements would be done via a quite different methodology, namely by direct weighing or other comparison with certain macroscopic bodies, for example monoisotopic silicon spheres, which have been calibrated as accurately as possible by counting the number of atoms in them and using those atoms' measured absolute masses, and/or by alternative indirect methods such as provided by the Watt balance*". And - Wignall explains - masses of atoms can presently be measured in terms of their de Broglie frequency with a precision of a few parts in $10^9$.

Having defined the mass to be a fundamental unit, the unit of energy follows very simply and exactly as through current *Système International d'Unités*: joule$[\mathrm{J}] = [\mathrm{kg} \cdot \mathrm{m}^2 \cdot \mathrm{s}^{-2}]$.

**References**

1. Einstein, A.: Annalen der Physik, **18** (1905) 639

2. Plank, M.: "Zur Dynamik bewegter Systeme", Berliner Sitzungsberichte (1907) 542

3. Ives, H.E.: "Derivation of the Mass-Energy Relation", Journal of the Optical Society of America **42** (1952) 540

4. Sachs, M.: "On the Meaning of E=mc$^2$", International Journal of Theoretical Physics **8** (1973) 377

5. Stachel, J., Torretti, R.: "Einstein's first derivation of mass-energy equivalence", American Journal of Physics **50** (1982) 760.

6. Jammer, M.: Concepts of Mass in Classical and Modern Physics. Harvard University Press (1961); Harper & Row (1964); Dover Publications (1997)

7. Jammer, M.: Concepts of Mass in Contemporary Physics and Philosophy. Princeton University Press (1999)

8. Pendse, C.G.: "A Note on the Definition and Determination of Mass in Newtonian Mechanics", Philosophical Magazine **24** (1937) 1012

9.. Narlikar, V.V.: "The Concept and Determination of Mass in Newtonian Mechanics", Philosophical Magazine **27** (1938) 33

10. McKinsey, J.C.C., Sugar, A.C., Suppes, P.: "Axiomatic Foundations of Classical Particle Mechanics", Journal of Rational Mechanics and Analysis **2** (1953) 253

11. Carnap, R.: An Introduction to the Philosophy of Science. New York (1966) 103

12. Koslow, A.: "Mach's Concept of Mass: Program and Definition", Synthese **18** (1968) 216

13. Goodinson, A., Luffman, B.L.: "On the Definition of Mass in Classical Physics", American Journal of Physics **53** (1985) 40

14. Kamlah, A.: "Zur Systematik der Massendefinitionen", Conceptus **22** (1988) 69

15. Schmidt, H. J.: "A Definition of Mass in Newton-Lagrange Mechanics", Philosophia Naturalis **30** (1993) 189

16. Sant'Anna, A.S.: "An Axiomatic Framework for Classical Particle Mechanics without Force", Philosophia Naturalis **33** (1996) 187

17. Zanchini, E., Barletta, A.: "A rigorous definition of mass in special Relativity", Il Nuovo Cimento B, **123** (2008) 153

18. Steck, D.J., Rioux, F.: "An elementary Development of Mass-Energy Equivalence", American Journal of Physics **51** (1983) 461

19. Feigenbaum, M.J., Mermin, N.D.: "$E=mc^2$", American Journal of Physics **56** (1988), 18

20. Rohrlich, F.: "An Elementary Derivation of $E=mc^2$", American Journal of Physics **58** (1990), 348

21. Srivastava, V.P.: "A Simple Derivation of $E=mc^2$", Physics Education **26** (1991) 214

22. Baierlein, R.: "Teaching $E=mc^2$", The physics Teacher **29** (1991) 170

23. Rindler, W.: Relativity: Special, General, and Cosmological. Oxford University Press - 2nd ed. (2006) 113

24. Landau, L D., Lifshitz, E.M.: The Classical Theory of Fields, 4$^{th}$ ed. Butterworth-Heinemann (1980) 25

25. Noether, E.: "Invariante Variationsprobleme". Nachr. D. König. Gesellsch. d. Wiss. zu Göttingen, Math-phys. (1918) 235.

26. Einstein, A.: "Prinzipielles zur allgemeinen Relativitätstheorie", Annalen der Physik **55**, 240 (1918)

27. Baker E.F.: American Journal of Physics **14** (1946) 309

28. Dicke R.H.: Science **129** (1959) 621

29. Higgs P.W.: Physics Letters **12** (1964) 132

30. Feynman, R.: QED: The Strange Theory of Light and Matter. Princeton University Press (1985).

31. de Broglie, L.: Philosophical Magazine 47 (1924) 446; Annales de Physique **3** (1925) 22

32. Planck, M.: Verh. D. Phys. Ges., 2 (1900) 202; ibid. **2**, 237

33. CODATA (Committee on Data for Science and Technology) 2007 values. They are based on all of the data available through 31 December 2006.

34. Wignall, J.W.G.: Metrologia **44** (2007) L19